\documentstyle[epsfig]{aipproc}
\begin{document}
\title{VHE $\gamma$-ray spectral properties\\
of the blazars Mrk~501 and Mrk~421 from CAT observations in 1997 and 1998}

\author{Fr\'ed\'eric Piron, for the CAT collaboration}
\address{Laboratoire de Physique Nucl\'eaire des Hautes Energies\\
Ecole Polytechnique, route de Saclay, 91128 Palaiseau Cedex, France}

\maketitle
\begin{abstract}
The Very High Energy (VHE) $\gamma$-ray emission of the BL~Lacert{\ae} objects Markarian~501 and
Markarian~421 has been observed by the C{\small AT} Imaging Atmospheric Cherenkov Telescope in
1997 and 1998. The spectrum extraction method is presented, and the spectral properties of
both sources are compared in different activity states.
Theoretical implications for jet astrophysics are discussed.
\end{abstract}

\section*{Spectral analysis method}
The C{\small AT} (Cherenkov Array at Th{\'e}mis) telescope
records Cherenkov flashes due to VHE atmospheric showers through its 17.8m$^2$ mirror.
The fine grain of its camera, combined with fast electronics, allows a relatively low $\gamma$-ray
detection threshold energy of $250\:\mathrm{GeV}$ (at Zenith) as well as an accurate analysis of the resulting images.
The experiment is fully described elsewhere~\cite{Barrau}. The analysis method is based on the fit of each
individual shower image with theoretical mean $\gamma$-ray images coming from a semi-analytical model~\cite{LeBohec}.
This permits the extraction of a $\gamma$-ray signal
from the background due to cosmic-ray induced atmospheric showers; the optimization of the event
selection using the Crab nebula data gives a hadron-rejection factor of $\sim$200 for a $\gamma$-ray efficiency
of $\sim$40\%. The fit also yields the energy of each shower in the $\gamma$ hypothesis, with a resolution of
$\sim$20\%, independent of energy, when restricting to showers with a small impact parameter ($<130\:\mathrm{m}$).
However, the spectrum reconstruction procedure presented below involves the exact energy-resolution function
$\Psi(E\rightarrow\widetilde{E},cos\theta)$~\footnote{Probability density, for fixed zenith angle $\theta$ and
{\it real} energy $E$, to get an {\it estimated} energy $\widetilde{E}$.}
in order to use the entire available statistics. These functions have been determined by detailed Monte-Carlo
simulations of the telescope response, as has the effective detection area ${\mathcal{A}}_{\mathrm{eff}}$,
which includes the effect of event-selection efficiency. The simulations have been checked and
calibrated on the basis of several observables by using muons rings and the nearly-pure $\gamma$-ray
signal from the highest flare of Mrk~501 in 1997~\cite{TeV99b}.\\
\indent With typical statistics of $\sim$1000 $\gamma$-ray events and a signal-to-background ratio of $0.4$
(as obtained on the Crab nebula), a spectrum can be determined with reasonable accuracy as follows.
First we define a set
$\{\Delta_{i_{e}}\}=\{[\widetilde{E}^{\mathrm{min}}_{i_{e}}, \widetilde{E}^{\mathrm{max}}_{i_{e}}]\}_{i_{e}=1,n_e}$ of
energy bins; since the threshold energy of the telescope increases with zenith angle, each of these is divided into a subset
$\{\Delta_{i_{e},i_{z}}\}=\{[\theta^{\mathrm{min}}_{i_{z}}, \theta^{\mathrm{max}}_{i_{z}}]\}_{i_{z}=1,n_z(i_{e})}$
of zenith angle bins~\footnote{The $i_{z}=1$ bin corresponds to the transit of the source.} with a width of
$\delta(\cos\theta)=0.05$. Then, for all {\small ON} and {\small OFF}-source runs~\footnote{Each run lasts
$\sim$30min, {\small OFF}-source runs being used to estimate the hadronic background.}, 
the number of events passing the selection cuts is determined within each $\Delta_{i_{e},i_{z}}$ 2D-bin. Finally,
assuming a given spectral shape, a maximum-likelihood estimation of the spectral parameters is performed,
which takes into account the effective detection area and the energy-resolution function. At this stage, two
hypotheses ${\mathcal{H}}^{\mathrm{pl}}$ and ${\mathcal{H}}^{\mathrm{cv}}$ are successively considered, which are
a simple power law ${\phi}^{\mathrm{pl}}(E) = {\phi}_0 E_{\mathrm{TeV}}^{-\gamma}$ and a curved shape
${\phi}^{\mathrm{cv}}(E) = {\phi}_0 E_{\mathrm{TeV}}^{-(\gamma + \beta\log_{10}\!E_{\mathrm{TeV}})}$, respectively.
Power laws often account for spectral properties in high energy astrophysics, at least because they often constitute
a good approximation when considered over a restricted energy range. On the other hand, a curved shape is suggested
by general considerations on emission processes within blazar jets (see the next
sections); the
latter parametrization, previously used by the Whipple group \cite{Samuelson}, corresponds to a
parabolic law in a $\log(\phi)$ {\it vs.} $\log (E)$ representation and allows simple comparisons.
The relevance of ${\mathcal{H}}^{\mathrm{pl}}$ with respect to ${\mathcal{H}}^{\mathrm{cv}}$ is estimated through
the likelihood ratio of the two hypotheses, which is defined as
$\lambda=-2\times\log(\frac{{\mathcal{L}}^{\mathrm{pl}}}{{\mathcal{L}}^{\mathrm{cv}}})$: it behaves
(asymptotically) like a $\chi^2$ with one d.o.f. and permits a search for the presence of
spectral curvature. {\it All spectral curves shown in the next sections come from the best set of fitted
parameters obtained by this procedure. It is also customary to superimpose on them ``experimental'' points
$\{{E}_{i_{e}}^{\mathrm{exp}}, {\phi}_{i_{e}}^{\mathrm{exp}}\}_{i_{e}=1,n_e}$, which are only indicative of the
statistics used in each $\Delta_{i_{e}}$ energy bin}. The definition of these points is the following: if we note
${\phi}^{\mathrm{best}}(E)$ as the best fitted spectrum under the hypothesis ${\phi}^{\mathrm{pl}}$ or ${\phi}^{\mathrm{cv}}$
according to $\lambda$,
$N_{i_{e},i_{z}}^{\mathrm{obs}}$ as the number of $\gamma$-ray events
observed in the $\Delta_{i_{e},i_{z}}$ 2D-bin, and ${\sigma}_{i_{e},i_{z}}$ as its error, and
$N_{i_{e},i_{z}}^{\mathrm{pred}} =
\displaystyle{\int}_{\widetilde{E}^{\mathrm{min}}_{i_{e}}}^{\widetilde{E}^{\mathrm{max}}_{i_{e}}}{\mathrm d}\widetilde{E}
{\int}_{0}^{\infty}{\mathrm d}E\;
{\phi}^{\mathrm{best}}(E)\times{\mathcal{A}}_{\mathrm{eff}}(E,\langle\cos\theta\rangle_{i_{z}})
\times\Psi(E\rightarrow\widetilde{E},\langle\cos\theta\rangle_{i_{z}})$ as the corresponding predicted number,
then ${E}_{i_{e}}^{\mathrm{exp}}$ is the unique energy within the bin for which
${\phi}^{\mathrm{best}}({E}_{i_{e}}^{\mathrm{exp}}) = {\langle{\phi}^{\mathrm{best}}\rangle}_{i_{e}}$
holds~\footnote{The average is taken over the bin according to the physical meaning of the flux,
i.e. by integrating over $E$ or $\log (E)$ in a $\mathrm{d}N/\mathrm{d}E$ or in a
$\nu F_\nu (\equiv E^2\times\mathrm{d}N/\mathrm{d}E$) representation, respectively.}, and
${\phi}_{i_{e}}^{\mathrm{exp}} = {\phi}^{\mathrm{best}}({E}_{i_{e}}^{\mathrm{exp}})\times 
\displaystyle\left[\sum_{i_{z}=1}^{n_z(i_{e})}\left(N_{i_{e},i_{z}}^{\mathrm{obs}}/N_{i_{e},i_{z}}^{\mathrm{pred}}\right)/
{\sigma}_{i_{e},i_{z}}^{2}\right] / 
\displaystyle\left[\sum_{i_{z}=1}^{n_z(i_{e})} 1/{\sigma}_{i_{e},i_{z}}^{2}\right]$.
It should be noted that this way of representing spectra cannot replace the likelihood method, which provides the
only relevant physical results, i.e. the values of the most probable spectral parameters as well as their errors
and covariance matrix.
{\it In particular, one should not use the $\{{E}_{i_{e}}^{\mathrm{exp}}, {\phi}_{i_{e}}^{\mathrm{exp}}\}$ points in any kind of
minimization, since the ${\sigma}_{i_{e},i_{z}}$ are correlated in a very complex manner.}
\section*{CAT observations of Mrk~501 and Mrk~421}
\subsection*{Spectral properties of Mrk~501 in 1997 and 1998}
\vspace{-.5cm}
\begin{figure}[h!]
\centerline{
\hbox{
\epsfig{file=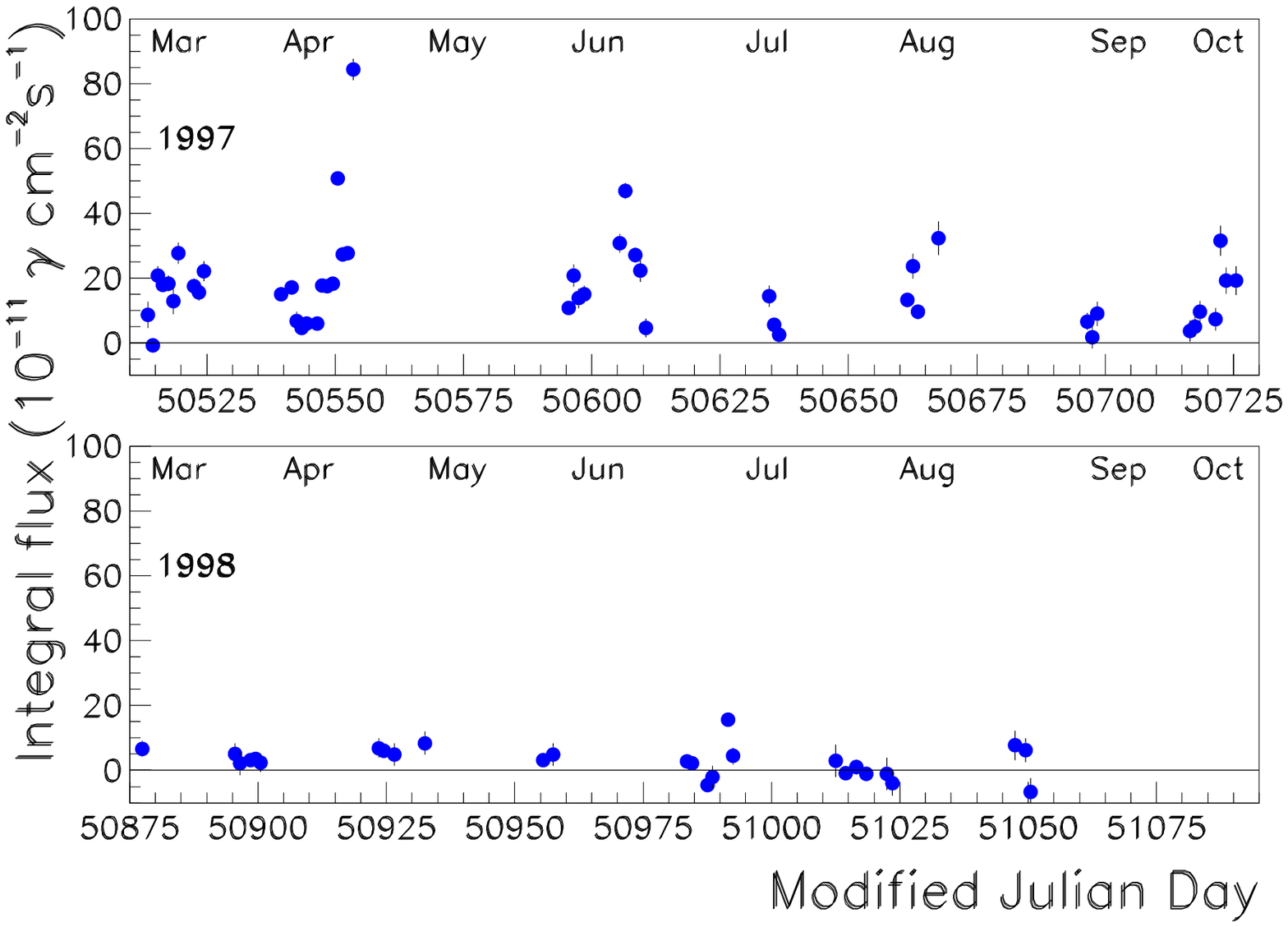,width=7cm,height=5cm,clip=
,bbllx=4pt,bblly=4pt,bburx=524pt,bbury=377pt}
\epsfig{file=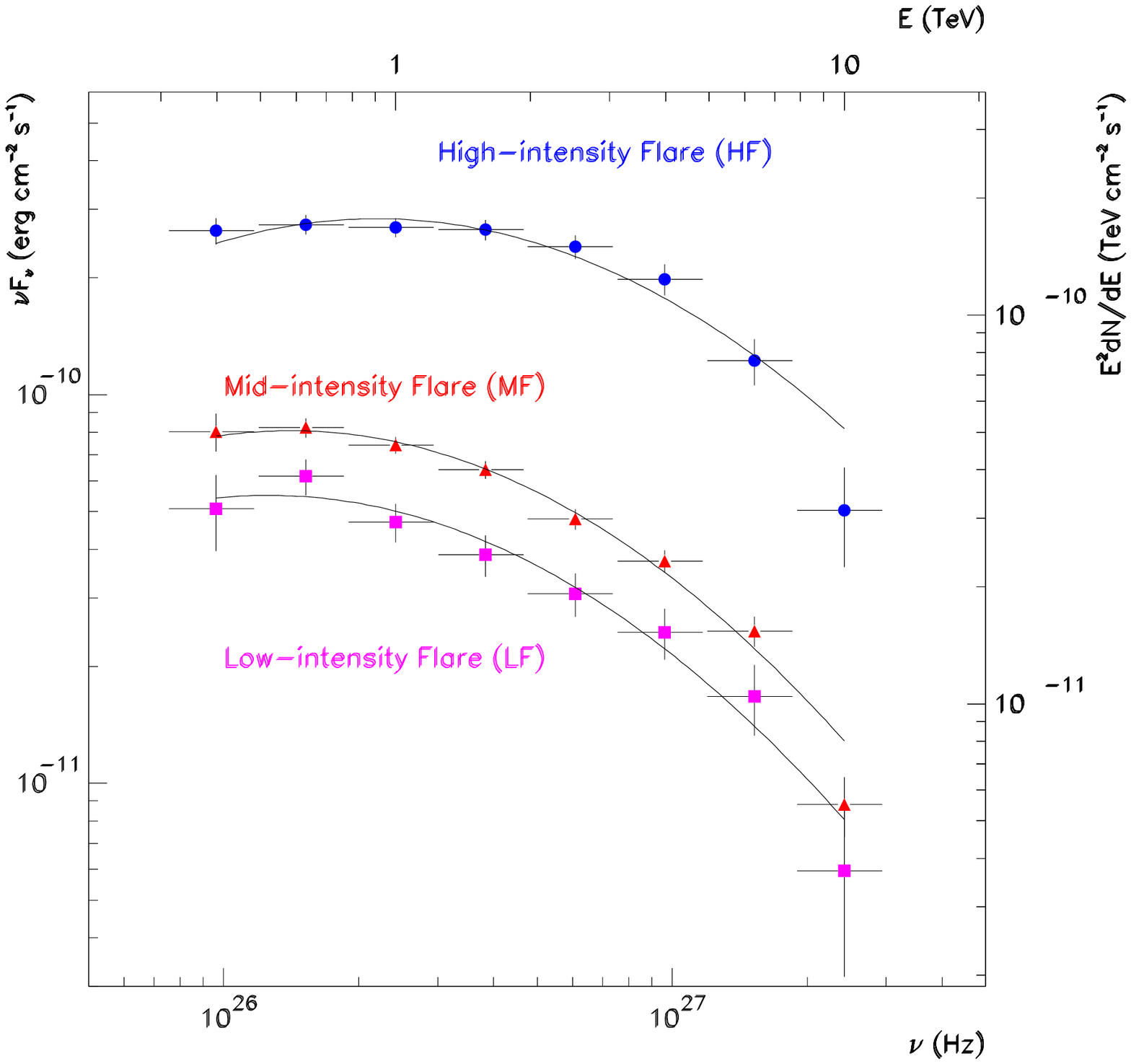,width=7cm,height=5cm,clip=
,bbllx=10pt,bblly=17pt,bburx=545pt,bbury=522pt}
}}
\caption{
{\it (a)} Mrk~501 nightly integral flux above $250\:\mathrm{GeV}$ in 1997 and 1998;
{\it (b)} Mrk~501 VHE SED between $330\:\mathrm{GeV}$ and $13\:\mathrm{TeV}$ for three independent data-subsets,
corresponding to different activity states in 1997 (see $[4]$ for details).
}
\label{501cl3nfn}
\end{figure}
Mrk~501 exhibited a remarkable series of flares during the whole year 1997, going
down to a much lower mean flux in 1998 (Fig.~\ref{501cl3nfn}a). The detailed spectral analysis of the 1997
data can be found in~\cite{Djannati}; here we extend it to the 1998 results.\\
\begin{figure}[b!]
\centerline{
\hbox{
\epsfig{file=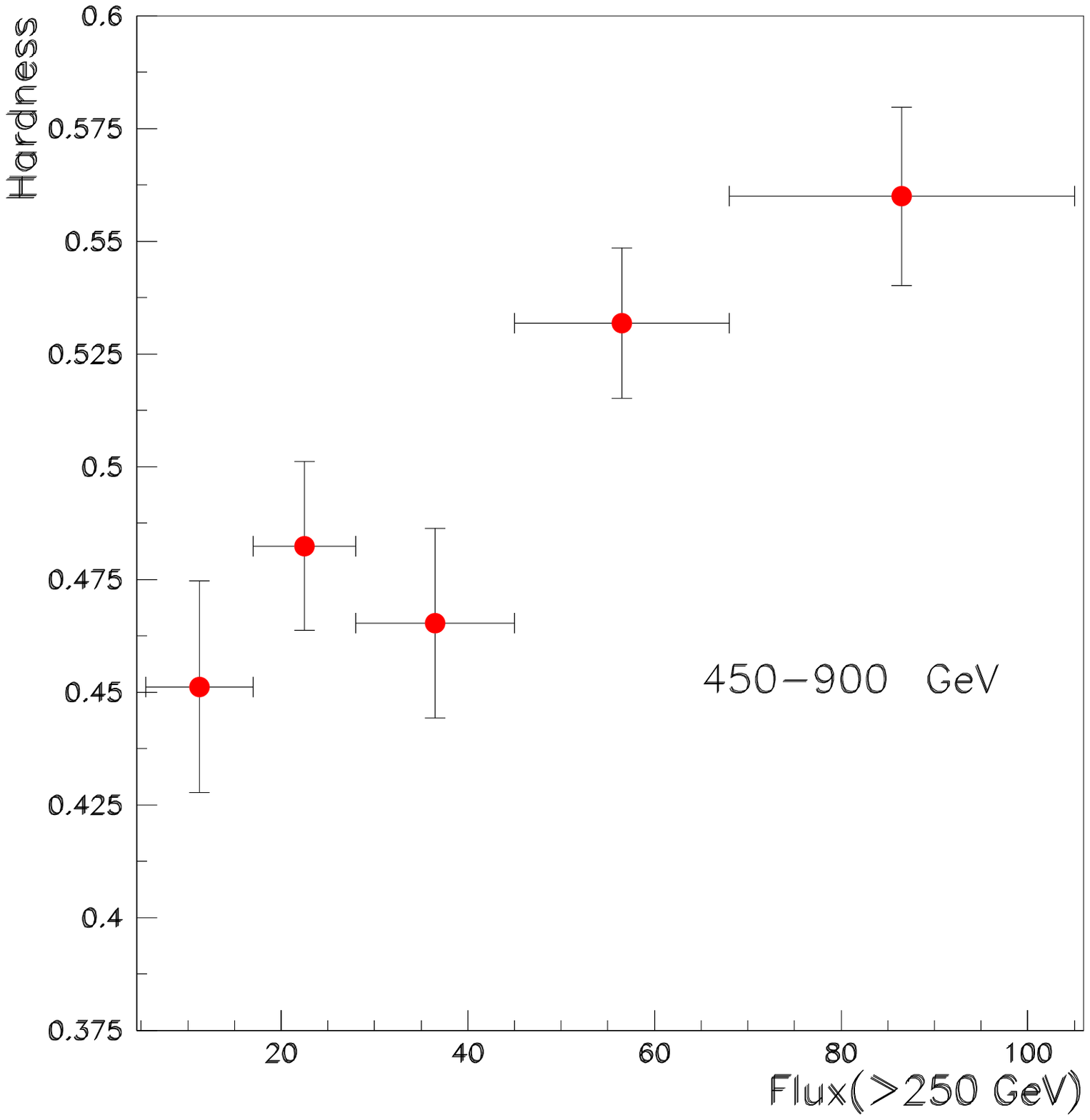,width=6.5cm,height=5cm,clip=
,bbllx=55pt,bblly=46pt,bburx=520pt,bbury=520pt}
\epsfig{file=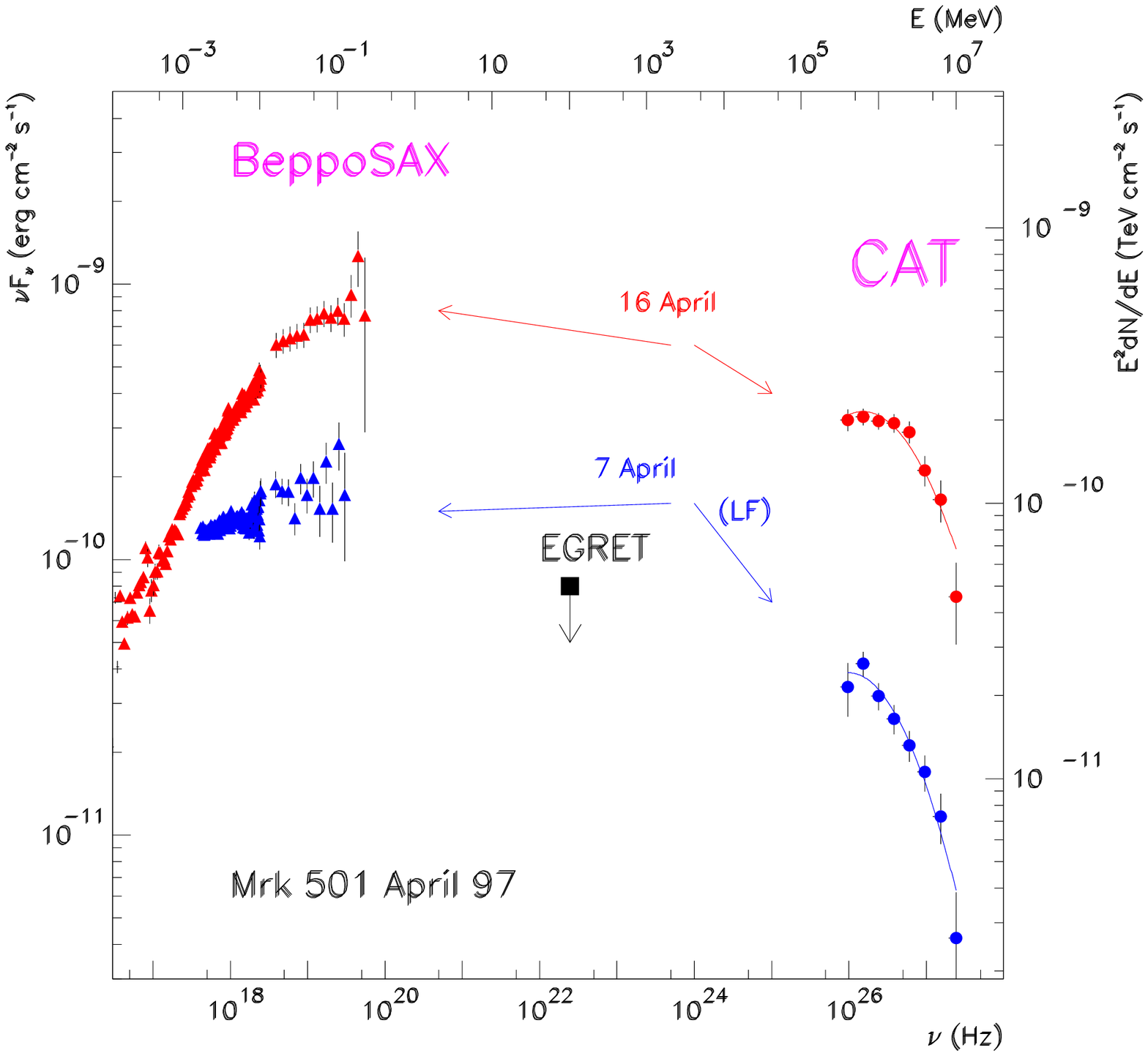,width=6.5cm,height=5cm,clip=
,bbllx=3pt,bblly=17pt,bburx=545pt,bbury=520pt}
}}
\caption{
{\it (a)} Hardness-ratio ($HR=\frac{N_{E>900\:\mathrm{GeV}}}{N_{E>450\:\mathrm{GeV}}}$) {\it vs.} source intensity
(${\Phi}_{>250\:\mathrm{GeV}}$ in units of $10^{-11}\:\mathrm{cm^{-2}s^{-1}}$);
{\it (b)} Mrk~501 X-ray and VHE spectra for April $7^{\mathrm{th}}$ and $16^{\mathrm{th}}$.
The EGRET upper limit corresponds to observations between April $9^{\mathrm{th}}$ and $15^{\mathrm{th}}$.
}
\label{501hrbeppo}
\end{figure}
\indent The VHE spectral energy distribution (SED) of Mrk~501, derived for different flaring-activity states in 1997
(Fig.~\ref{501cl3nfn}b), shows a significant curvature which is now well confirmed by different ground-based
experiments~\cite{Aharonian1,Krennrich}. The peak $\gamma$-energy is found to lie just above the C{\small AT}
threshold, and it seems to shift towards higher energies as the flux increases. To
check this spectral variability by a more robust method, the hardness ratio has been computed
for five different-level intensities:
the correlation observed in Fig.~\ref{501hrbeppo}a confirms the hardening of the VHE SED during flaring periods.
Fig.~\ref{501hrbeppo}b shows the broad-band SED of Mrk~501 for two flaring dates in April 1997: it
exhibits the two-bump structure typical of blazars, with a dip indicated by the contemporary E{\small GRET} upper-limit point
in the GeV energy range~\cite{Samuelson}. The obvious correlation of the X-ray emission from BeppoS{\small AX}
data~\cite{Pian} with TeV emission strongly suggests that the same particle
population is responsible for emission in both energy-ranges, i.e. it supports the picture given by leptonic
models~\cite{Ghisellini}, in which an energetic electron beam propagating in the magnetized plasma jet produces
X-rays through synchrotron radiation as well as VHE $\gamma$-rays through inverse Compton scattering of low-energy
photons.
\begin{figure}[h!]
\centerline{
\hbox{
\epsfig{file=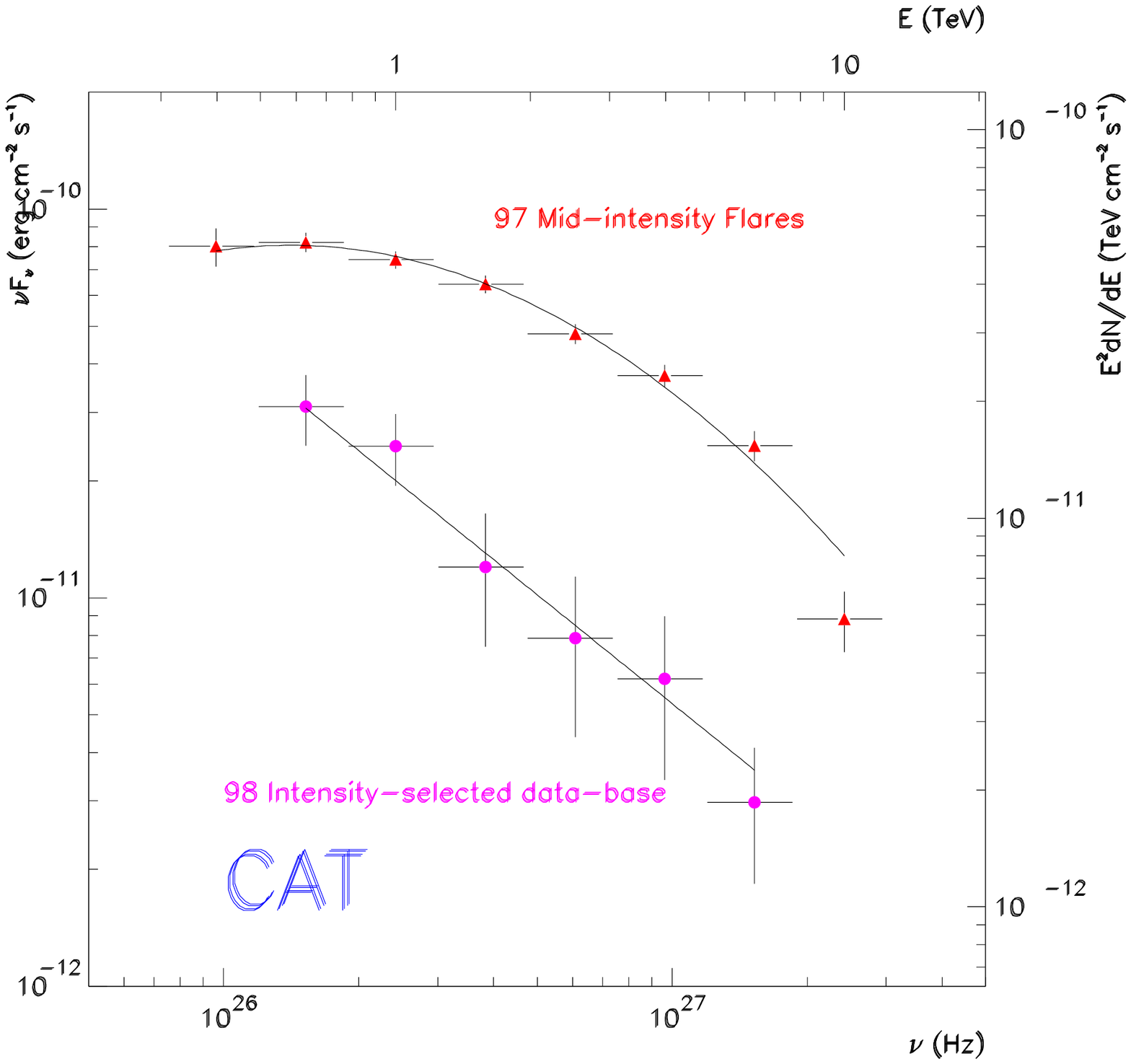,width=6.5cm,height=5cm,clip=
,bbllx=11pt,bblly=18pt,bburx=545pt,bbury=522pt}
\epsfig{file=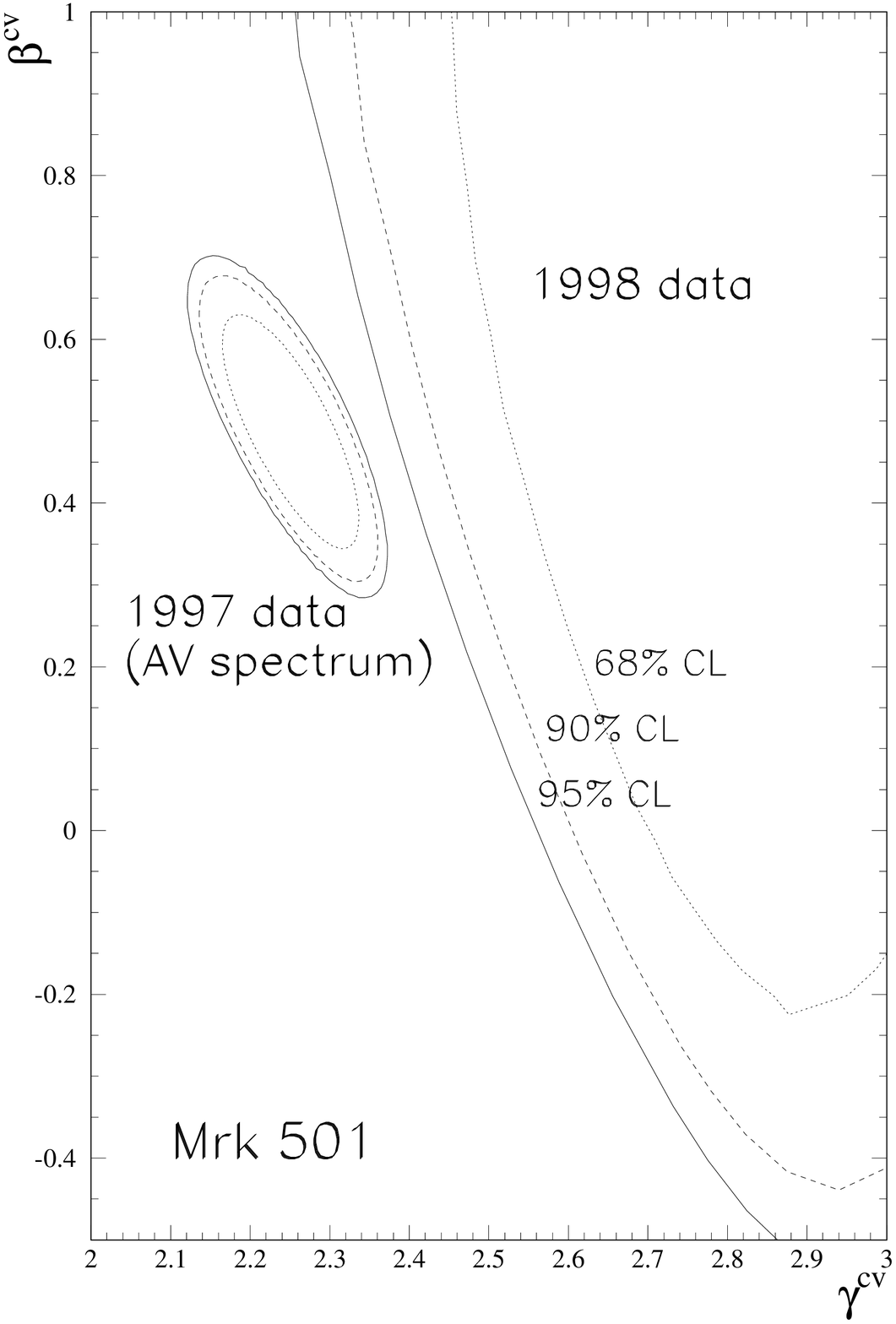,width=6.5cm,height=5cm,clip=
,bbllx=30pt,bblly=35pt,bburx=540pt,bbury=800pt}
}}
\caption{
{\it (a)} Mrk~501 VHE SED from $330\:\mathrm{GeV}$ to $5.2\:\mathrm{TeV}$ in 1998,
as compared to the MF data subset
in 1997 (see Table~\ref{tabsummary});
{\it (b)} $68$\%, $90$\% and $95$\% CL contours in the $\{\gamma^{\mathrm{cv}}, \beta^{\mathrm{cv}}\}$ plane for the 1997
AV and 1998 data sets (see Table~\ref{tabsummary});
each contour is obtained by projecting the 3-dimension ellipsoid along the $\phi_0^{\mathrm{cv}}$ axis.
}
\label{501sp98}
\end{figure}\\
\indent Unlike the 1997 spectra, the spectrum of Mrk~501 in 1998 does not show any
curvature (Fig.~\ref{501sp98}a),
indicating a VHE peak energy well below the C{\small AT} threshold at that time.
Fig.~\ref{501sp98}b confirms in particular that, in spite of larger statistical errors, the
1998 data cannot be fitted by the average spectrum shape found in 1997 (AV data set, see Table~\ref{tabsummary}),
as would be expected in the absence of any spectral variability. With a mean flux much lower than that of 1997,
the power-law shape found in 1998 is therefore consistent with the previously suggested scenario, in which the
peak $\gamma$-energy seems to be correlated with the VHE flux.
\subsection*{Mrk~421's spectrum in 1998}
\begin{figure}[t!]
\centerline{
\hbox{
\epsfig{file=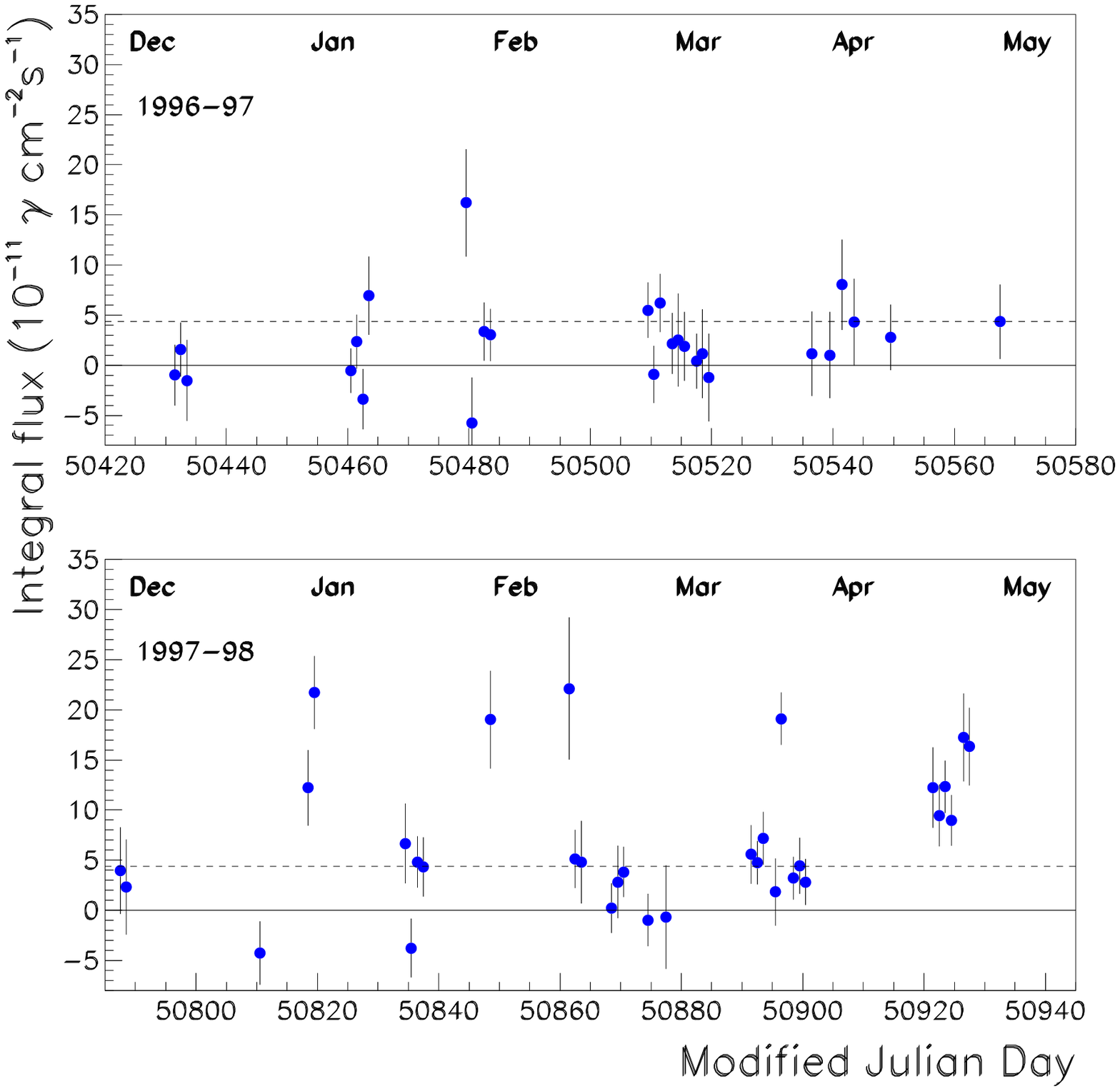,width=6.5cm,height=5cm,clip=
,bbllx=3pt,bblly=4pt,bburx=535pt,bbury=520pt}
\epsfig{file=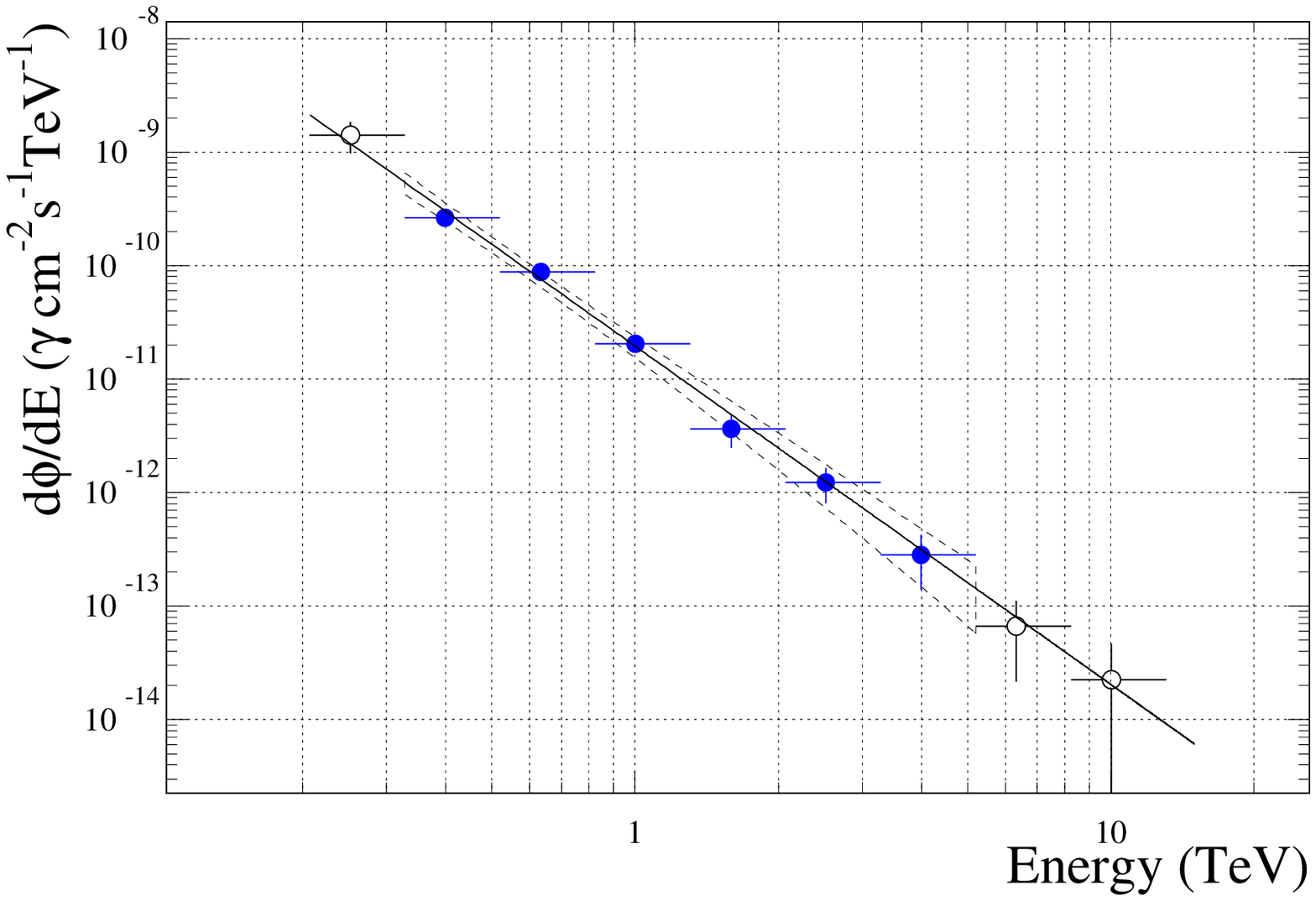,width=6.5cm,height=5cm,clip=
,bbllx=15pt,bblly=4pt,bburx=545pt,bbury=363pt}
}}
\caption{
{\it (a)} Mrk~421 nightly integral flux above $250\:\mathrm{GeV}$
between December 1996 and May 1998. The dashed line represents the mean flux over the two years;
{\it (b)} Differential flux of Mrk~421 for the flaring periods in 1998. Only the energy bins shown with
fullfilled circles (from $330\:\mathrm{GeV}$ to $5.2\:\mathrm{TeV}$) were used in the likelihood method (see text).
}
\label{421clsp}
\end{figure}
As reported in~\cite{PironICRC}, Mrk~421 is the second extragalactic source detected by C{\small AT}:
almost quiet in 1996-97,
the source showed small bursts in 1997-98, together with a higher mean activity (Fig.~\ref{421clsp}a).
The energy spectrum derived for the 1998 flaring periods
is well represented by a simple power law (Fig.~\ref{421clsp}b) with a differential spectral index
$\gamma = 2.96 \pm \mathrm{ 0.13^{stat} \pm 0.05^{syst}}$~\cite{PironICRC}. This agrees with a recent result
of the H{\small EGRA} group concerning the 1997-98 period~\cite{Aharonian2}.
It is however in contrast with the former behaviour of the source at the time of the 1995-96 flaring period,
as observed by the Whipple group, who found $\gamma = 2.54 \pm \mathrm{0.03^{stat} \pm 0.10^{sys}}$~\cite{Krennrich}.
Thus, the higher value of the differential spectral index found in this work could come from a correlation between
the intensity level and the spectral hardness, like that observed for Mrk~501 in 1997
(see~\cite{Djannati} and above). A similar evidence for Mrk~421 was in fact suggested in \cite{Zweerink} on the basis
of the Whipple 1995-96 low-flux data.\\
\indent In the framework of leptonic models~\cite{Ghisellini}, which succesfully explain the
Mrk~501 broad-band SED in 1997, the absence of any obvious spectral curvature reported by
all experiments implies in any case that the peak energy of the inverse-Compton contribution to Mrk~421's spectrum
is significantly lower than the C{\small AT} detection threshold, as for Mrk~501 in 1998.
This is not surprising since the corresponding synchrotron peak is lower than that of Mrk~501 in 1997, and
since we have seen that leptonic models predict a strong correlation between X-rays and $\gamma$-rays. 
In fact, such a correlation was directly observed on Mrk~421 in Spring 1998, during
a coordinated observation campaign involving ground-based Cherenkov imaging telescopes (Whipple,
H{\small EGRA}, and C{\small AT}) and the A{\small SCA} X-ray satellite\cite{Takahashi99}.
\begin{table*}[t!]
\caption{Best fitted spectral parameters of Mrk~501 and Mrk~421 in 1997 and 1998
from the ${\mathcal{H}}^{\mathrm{pl}}$ and
${\mathcal{H}}^{\mathrm{cv}}$ shape assumptions. The data-subsets of Mrk~501 in 1997 are defined in Fig.~\ref{501cl3nfn}b,
while the average set (AV) contains the data of the whole year.
Fluxes are given in units of $10^{-11}\:\mathrm{cm^{-2}s^{-1}TeV^{-1}}$.
If, according to $\lambda$, the ${\mathcal{H}}^{\mathrm{cv}}$ hypothesis is favored, then the last two columns gives the
curvature term $\beta$ and the peak-emission energy $E^{\mathrm{peak}}_{\mathrm{GeV}}=10^\frac{2-\gamma}{2\beta}$.
If not, $\beta$ is quoted between brackets.}
\label{tabsummary}
\begin{tabular}{rrccccc}
Data set&T~\tablenote{total time (h) ON-source.}&$\lambda$&$\phi_0$&$\gamma$&$\beta$&$E^{\mathrm{peak}}_{\mathrm{GeV}}$\\
&&&&&&\\
\hline
\hline
\noalign{\smallskip}
\noalign{\smallskip}
Mrk~501'97 LF&$13.6$&$10.7$&$\;\:3.13\pm0.19$&$2.32\pm0.09$&$0.41\pm0.17$&$410\pm201$\\
MF&$40.5$&$47.1$&$\;\:4.72\pm0.14$&$2.25\pm0.05$&$0.52\pm0.08$&$583\pm104$\\
HF&$3.1$&$29.1$&$17.60\pm0.61$&$2.07\pm0.04$&$0.45\pm0.09$&$840\pm108$\\
AV&$57.2$&$61.5$&$\;\:5.19\pm0.13$&$2.24\pm0.04$&$0.50\pm0.07$&$578\pm\;\;98$\\
\noalign{\smallskip}
\hline
\noalign{\smallskip}
Mrk~501~'98&$3.3$&$0.09$&$\;\:1.25\pm0.16$&$2.97\pm0.20$&$(0.21\pm0.73)$&--\\
\noalign{\smallskip}
\hline
\noalign{\smallskip}
Mrk~421~'98&$5.1$&$0.34$&$\;\:1.96\pm0.20$&$2.96\pm0.13$&$(0.28\pm0.49)$&--\\
\noalign{\smallskip}
\end{tabular}
\end{table*}
\section*{Conclusion}
Since Mrk~501 and Mrk~421 lie at the the same redshift ($\sim$0.03), spectral differences between them must
be intrinsic and not due, in particular, to absorption by the diffuse infrared background radiation. This allows
direct and relevant comparison of their spectral properties. The correlation between X-ray and $\gamma$-ray
emissions is now proven for both sources, supporting the simple and most natural scenario given by leptonic
models~\cite{Ghisellini} for the origin of blazar TeV flares, in which a single leptonic population is injected into
the radio jets and produces correlated X-ray synchrotron and VHE $\gamma$-ray inverse Compton radiations.
C{\small AT} observations complete this picture with some evidence of spectral variability of Mrk~501 in 1997,
suggesting that this leptonic population is responsible for the hardening of the entire high-energy part of the
electromagnetic spectrum during flares. To date, the peak $\gamma$-energy of Mrk~421 has always remained well below the
C{\small AT} threshold, precluding a more accurate spectral study. Therefore, testing VHE spectral variability as a general
feature of blazars requires more multiwavelength observations with a large dynamic range in intensity.
In any case, the comparison of the spectral properties of Mrk~501 and Mrk~421 in 1997 and 1998 fits the
picture in which the VHE peak $\gamma$-energy shifts with increasing VHE flux: as shown in
Table~\ref{tabsummary}, spectral curvature seems to be characteristic of high flaring-activity states
(i.e. Mrk~501 in 1997), whereas low-activity spectra (i.e. Mrk~501 and Mrk~421 in 1998) are always compatible with
pure power-laws.

\end{document}